\documentstyle[preprint,aps,epsf,floats,tighten]{revtex}
\begin{document}
\draft
\title{Intersecting branes with an arbitrary excess angle}
\author{Tibor Torma\thanks{E\_mail address: kakukk@physics.utoronto.ca}}
\address{Department of Physics, University of Toronto, 
Toronto, Ontario, Canada M5S~1A7}
\date{\today}
\maketitle
\begin{abstract}
We find all possible static embeddings of a 4-brane in any dimension-6 spacetime with
4-dimensional Poincar\'e symmetry and a negative cosmological constant, subject to
orbifolding across the brane. Our new solutions allow intersecting branes at an
angle determined by a new dynamical parameter.  A collection of branes intersecting
in one 3-brane allows an arbitrary excess angle that can be related to a vacuum
density along the intersection. The 3-brane must be stabilized by additional
fine-tuned interactions. We also note that localization of gravity is tied to the
approximate fine tuning of the brane tensions.
\end{abstract}
\pacs{11.27.+d,04.50.+h,11.25.Mj}
\narrowtext

\section{Introduction}

In the seminal paper~\cite{RS2} gravity lives in five-dimensional anti~de-Sitter
($AdS_5$) spacetime, infinite but strongly curved. Due to the presence of one
graviton bound state, related to the finite proper size of ``bulk" spacetime,
gravity is effectively localized to a 4-dimensional ``brane". This mechanism however
works only on codimension~1 branes but could be indirectly
generalized~\cite{Intersect1} by placing $D-4$ codimension~1 branes intersecting in
one 3-brane (our world). The brane tension of each of these 4-branes localizes
gravity in the respective dimension. Models have been built and their cosmological
consequences analyzed in~\cite{IntersectAll}.

This geometric setup has also been used to address the cosmological constant.
Static embeddings of 3-branes in $AdS_5$ spacetimes were found~\cite{KaloperBent}
where the induced metric is dS, Minkowski or AdS; the curvature of the brane could
be made much smaller than the brane tension. The same solution, now for a 4-brane in
6~dimensions, was rediscovered in conformally flat form in~\cite{ANelson}. It was
observed in this paper that it is possible to embed a static 4-brane, plane in
conformal coordinates, at any angle $\phi$ determined by the brane tension,
$\sin\phi=\sigma/\sigma_c$, where $\sigma_c$ is a critical value for the brane
tension. A vanishing observed cosmological constant (i.e. flat induced metric) is
produced when $\sigma=\sigma_c$. Cutting out wedges of the bulk $AdS_5$ with some
angle $\phi$ and pasting together $n$ copies can accommodate an excess angle
$n\phi-2\pi$. This must match the 3-brane vacuum energy in the intersection.
The wedges are stabilized by imposing orbifold boundary conditions on each
4-brane. Whereas in the original Randall-Sundrum models~\cite{RS2,RS1} the 3-brane
tension is fine tuned to the bulk cosmological constant, here it must be fine tuned
to match the excess angle, a function of the 4-brane tensions.

A similar setup with different bulk cosmological constants in each wedge is
considered in~\cite{Csaki}, although without the introduction of an excess angle and
correspondingly, without a 3-brane tension. It has been speculated in this reference
that an additional parameter of the solution could resolve the one remaining fine
tuning. What we need is a flat direction; then a new dynamical mechanism could
stabilize the cosmological constant at zero value. Any additional vacuum energy
would result in the readjustment of the brane angles and the additional parameter.

In this paper we find such an additional parameter. Using Gaussian normal
coordinates (GNC) of a 4-brane, we write down all vacuum solutions in the bulk (with
a negative cosmological constant), satisfying the Ansatz with signature
$+++++-$\footnote{Allowing more than one timelike coordinate some 
restrictions on the brane tension disappear.}
\begin{equation}\label{eq:ansatz}
ds^2=dw^2+B(w,y)dy^2+A(w,y)\eta_{\mu\nu}dx^\mu dx^\nu.
\end{equation}
The 4-brane is located at $w=0$. The stability of the brane with orbifold boundary
conditions (i.e. $w\leftrightarrow-w$ identified while $y,x^\mu$ are fixed) comes
down to the Israel conditions~\cite{Israel} on the extrinsic curvature,
$K_{ab}=-\frac{\sigma}{2} h_{ab}$, where $\sigma$ is the 4-brane tension in
appropriate units ($h_{ab}$ is the extrinsic curvature of the brane).

There is considerable freedom in transforming the coordinates, still within the
Ansatz of Eq.~(\ref{eq:ansatz}). In addition, the same bulk metric is found starting
with brane with different tensions, embedded at various angles. After eliminating
all these ambiguities we are left with four different solutions (in each case the
$y$ coordinate is chosen to render the induced metric conformally flat):
\begin{itemize}
\item{A solution $A$ in which only a supercritical tension brane can be embedded,
$\sigma>\sigma_c$, where the induced metric has two curvature singularities at
finite $y\rightarrow\pm y_*$ (these are also at finite proper distance)}
\end{itemize}
and three solutions where only $\sigma<\sigma_c$ branes can be embedded, namely
\begin{itemize}
\item{A solution $B$ where the induced metric has a curvature singularity at
finite proper and coordinate) distance $y\rightarrow-y_*$ and an  $AdS$ horizon at
$y\rightarrow y_+$.}
\item{A solution $C$ with $AdS$ horizon at both $y\rightarrow\pm y_+$.
The $w\rightarrow\infty$ limit is also asymptotically AdS.}
\item{A solution $D$ where both the bulk and the induced metric are AdS. This is the
solution found previously~\cite{KaloperBent,ANelson,Csaki}.}
\end{itemize}
(We do not consider the fine tuned case, $\sigma=\sigma_c$.)

All these solutions are vacuum solutions with a negative cosmological constant in the bulk,
i.e. solutions with constant scalar curvature. With the exception of $D$, however,
they are not maximally symmetric. The induced metric is not even a vacuum solution.
It is worth noting that the corresponding embedding of 3-branes in 5 dimensional
spacetimes can be similarly found, but the new solutions have little value because
the induced 3-brane metric does not resemble our world at all. 

All our solutions have four-dimensional Poincar\'e invariance and can be trivially
generalized to any maximally symmetric 4-dimensional metric replacing $\eta_{\mu\nu}$
in Eq.~(\ref{eq:ansatz}). Eventually the function $A(w,y)$ is the same in that case,
only $B(w,y)$ is changed. The solutions have a complicated mathematical form and are
expressed in an implicit form as solutions of algebraic equations involving special
functions.

The usual picture of localization of gravity to a codimension one brane acquires
some peculiar features when the brane tension differs from the fine tuned value. The
warp factor that effectively compactifies the bulk decreases up to a distance
${\cal O}\left(\frac{\sigma}{\sigma_c-\sigma}\,l_{AdS}\right)$ in the extra
dimension, but further away from the brane it starts to grow exponentially. The
resulting ``tail" of the graviton wave function eventually delocalizes brane
gravity. This effect is absent for exact fine tuning and {\it is also relevant for
3-branes embedded in a 5 dimensional bulk}, e.g. for the models
in~\cite{KaloperBent,ANelson,Csaki}.  Models in which the brane tensions are far from
their critical value do not seem to have brane gravity at all. We discuss in what
extent this can be considered a coordinate artifact in Sec.~\ref{subsec:localiz}.

At any point along the brane, another brane can be inserted at an angle determined
by its tension. In coordinates where the first brane has a conformally flat induced
metric, the second brane is curved. The (covariant) angle between the branes is
$\phi=\pi\pm 2\arcsin (\sigma/V)$. The quantity $V$ is independent of the tension of
the second brane but changes as the intersection point is moved along the first
brane ($\pm$ corresponds to two solutions and tells if $A$ increases or decreases as
one moves away from the intersection along the brane). For example, close to the
curvature singularity $V\rightarrow\infty$. Consequently, when several such wedges
are sewn together in the same point, all branes will have the same parameter $V$.
Now the induced metric on each brane is determined only by this $V$ (and the choice
of the $\pm$ sign), so that the first and last branes can always be identified in
order to build a manifold around the intersection. The orbifolding condition only
requires, for global consistency, that every second brane around the intersection
has the same tension (and $\pm$ sign).

The new parameter $V$ tells what part of the full 6-dimensional spacetime is cut
away. Its emergence is due to the fact that the solution is not maximally symmetric.
In the case of and $AdS_6$ solution, when the intersection point is shifted, the
change in the metric can be compensated by a coordinate transformation, i.e. all
points are physically equivalent. In our spacetime this is not so and $V$
parameterizes this difference. Because the angle of the wedges depends on $V$, so
does the total excess angle. The topological requirement that a vacuum density
produces an excess angle can be satisfied without fine tuning: the gravitational
equations simply set the brane angles and the parameter $V$ to the consistent
value.

When a brane tension is put on the 3-brane, another condition
arises~\cite{Chodos,Luty} which has not been discussed in~\cite{ANelson,Csaki}. A
positive 3-brane tension tries to minimize the 4-volume, which corresponds to
minimizing the value of $A(w,y)$ at the intersection. There is such a minimum only in
case~$C$. Now at the minimum the value of $V$ is zero, so we loose the
additional freedom we had when we only needed to stabilize the 4-branes. For
purposes of illustration we consider the following unusual but
consistent geometry. We take two branes (to be identified), of finite length. At
each end the two are joined and the covariant intersection angle is taken to be more
than $2\pi$. Upon identifying the two branes we find an asymptotically $AdS_6$
spacetime which satisfies the gravitational equations except the minimizing of
$A(w,y)$. Introducing some additional repulsive interaction between the two
3-branes, one can keep them several (6-dimensional) Plank lengths apart to ensure
the cancellation of the 4-dimensional cosmological constant. This requires a fine
tuning of the new coupling constants.

The paper is organized as follows. In Sec.~\ref{sec:solutions} we describe all
possible embeddings of a 4-brane with 4-dimensional Poincar\'e invariance and detail
our observations on the localization of gravity to the brane. Sec.~\ref{sec:branes}
shows how intersecting branes can be embedded and explains the emerging additional
parameter. In Sec.~\ref{sec:model} we show that these joining branes can be sewn
together in a global manifold and illustrate on a simple model why the additional
interaction strengths need to be fine tuned.

\section{Static 4-branes in 6 dimensions}\label{sec:solutions}

As explained in the introduction, we are looking for solutions to the
six-dimensional Einstein equations:
\begin{equation}\label{eq:einstein}
\overline{G}_{ab}\equiv R_{ab}-\frac{R}{2}g_{ab}-g_{ab}=0.
\end{equation}
These equations follow from the action with a number of branes included,
\begin{equation}\label{eq:action}
S[g]=\int_{bulk}(\Lambda_6-2M_6^4 R[g])+\int_{4-brane(s)}\Lambda_5,
\end{equation}
while the units are so chosen that the bulk cosmological constant is set to~$-1$:
\begin{equation}
\frac{\Lambda_6}{4M_6^4}\rightarrow-1.
\end{equation}

We are looking at metrics that satisfy the Ansatz of Eq.~(\ref{eq:ansatz}) and the
Israel conditions~\cite{Israel} at $w=0$,
\begin{equation}\label{eq:israeli}
K_{ab}=-\frac{\sigma}{2}h_{ab}.
\mbox{\ \ \ \ with\ \ \ \ }\sigma=\frac{\Lambda_5}{8M_6^3}
\end{equation}
These conditions ensure that keeping the $w\geq0$ half plane and pasting back on
another copy with $w\leftrightarrow-w$ identified can accommodate a 4-brane with
tension~$\sigma$. Note that the critical (``fine tuned") value of the 4-brane tension
is at $\sigma_c=\sqrt{\frac{2}{5}}$ in our notation.

A direct calculation shows that the Einstein equations in the bulk become
\begin{eqnarray}
0&=&\label{eq1}
\left(4\frac{A_{yy}}{A}+\frac{A_{y}^2}{A^2}-2\frac{A_{y}}{A}\frac{B_{y}}{B}\right) +
\left(3\frac{A_{w}^2}{A^2}+2\frac{A_{w}}{A}\frac{B_{w}}{B}-2\right) B\\ 
0&=&\label{eq2}
\left(6\frac{A_{yy}}{A}-3\frac{A_{y}}{A}\frac{B_{y}}{B}\right) +
\left(6\frac{A_{ww}}{A}+3\frac{A_{w}}{A}\frac{B_{w}}{B}+2\frac{B_{ww}}{B}-
\frac{B_w^2}{B^2}-4\right)B\\
 0&=&\label{eq3}
3\frac{A_y^2}{A^2} -
\left(2-\frac{A_w^2}{A^2}-4\frac{A_{ww}}{A}\right) B\\ 
0&=&\label{eq4}
\frac{A_yA_w}{A^2}-2\frac{A_{yw}}{A}+\frac{A_w}{A}\frac{B_w}{B}
\end{eqnarray}
while the Israel conditions are satisfied at $w=0$,
\begin{equation}
\frac{A_w}{A}=-\sigma\mbox{\ \ \ \ and\ \ \ \ }\frac{B_w}{B}=-\sigma.
\end{equation}

In the following we use the notation~$A(w,y)=e^{a(w,y)}$. Observe from
Eq.~(\ref{eq3}) that we either have $a_y=0$ and $2-5a_w^2-4a_{ww}=0$ simultaneously,
or $a_y\neq0$ and $2-5a_w^2-4a_{ww}>0$. By calculating the curvature tensor element
of ${R_{wtt}}^w$ we observe that a necessary condition to have a maximally symmetric
bulk solution\footnote{That is, one satisfying
\(R_{abcd}=\frac{R}{30}\left(g_{ac}g_{bd}-g_{ad}g_{bc}\right)\).} is
$2-5a_w^2-10a_{ww}\equiv0$.

In order to solve these equations in the $a_y\neq0$ case we first express $B$ from
Eq.~(\ref{eq3}),
\begin{equation}\label{eq:3'}
B=\frac{3a_y^2}{2-5a_w^2-4a_{ww}},
\end{equation}
and substitute this back into Eq.~(\ref{eq4}). The result contains only
$w$-derivatives and can be solved as an ordinary differential equation for $a_w$
along the $y=const$ lines, which by construction are the brane orthogonal geodesics
with affine parameter $w$:
\begin{equation}\label{eq:4'}
\partial_w\frac{\left(2-5a_w^2-10a_{ww}\right)^2}{\left(2-5a_w^2-4a_{ww}\right)^5}=0.
\end{equation}
This is a first order differential equation on $a_w$ with boundary condition
$a_w(w=0)=-\sigma$, so that once the $w$-independent function
\(c_1(y)={\left(2-5a_w^2-10a_{ww}\right)^2}/{\left(2-5a_w^2-4a_{ww}\right)^5}\) is
specified it has one unique solution. It is not hard to see that Eq.~(\ref{eq2}) is
a consequence of Eqs.~(\ref{eq3},\ref{eq:4'}). The boundary condition on $B_w$
trivially follows from~Eq.~(\ref{eq4}).  Now Eq.~(\ref{eq1}) relates the function
$c_1(y)$ to $a_0(y)\equiv a(0,y)$. Substituting $B$ from Eq.~(\ref{eq3}) into
Eq.~(\ref{eq1}) we find
\begin{equation}\label{eq:c0}
0=\partial_y \left(A\sqrt[5]{2-5a_w^2-10a_{ww}}\right)
\equiv\partial_y \left(\left[A\left(2-5a_w^2-4a_{ww}\right)\right]^5c_1(y)\right).
\end{equation}
The term in the last parenthesis,
\(e^{5c_0}\equiv\left[A\left(2-5a_w^2-4a_{ww}\right)\right]^5c_1(y)\), is
independent of $w$, as can be checked by showing that its $w$-derivative is
proportional to Eq.~(\ref{eq:4'}). Then Eq.~(\ref{eq1}) is equivalent to
$c_0=const$. We can calculate, with $\eta=sign\left(2-5a_w^2-10a_{ww}\right)=\pm1$,
\begin{equation}\label{eq:c1}
c_1(y)=\frac{\exp5[c_0-a_0(y)]}{\left\{\frac{3}{5}(2-5\sigma^2)+
\frac{2}{5}\eta\exp\frac{5}{2}[c_0-a_0(y)]\right\}^5}.
\end{equation}

We have found all stable embeddings of a 4-brane with tension~$\sigma$ satisfying
the Ansatz in terms of the arbitrary constant $c_0$, the sign $\eta$ and an
arbitrary function $a_0(y)$. The latter specifies the factor $A$ in the induced
metric on the brane.

The unique solution of Eq.~(\ref{eq:4'}) can be written in an implicit form as
follows. First introduce the quantity
\mbox{$v(y,w)=\frac{3}{5}\frac{2-5a_w^2}{2-5a_w^2-4a_{ww}}$},
which can be found from solving the fifth order equation
$(2-5\tau^2)^3\overline{c}_1 (y)=v^3(v-1)^2$ in terms of $\tau\equiv a_w$ and
$\overline{c}_1 (y)\equiv\frac{4\times27}{5^5}c_1(y)$; this is actually a
rewriting of the definition of $c_1(y)$. Multiple solutions of this equation will
lead to the different solutions mentioned in the introduction. Using the result we
calculate the integral
\begin{equation}
w=\int_{-\sigma}^{a_w(w,y)} \frac{d\tau}{r(\tau,y)} \mbox{\ \ \ \ with\ \ \ \ }
r(\tau,y)\equiv a_{ww}=\frac{3}{20}(2-5\tau^2)\left(\frac{5}{3}-\frac{1}{v}\right),
\end{equation}
which determines $a_w$ as a function of $w,y$. Integrating it,
\(a(w,y)=a_0(y)+\int_0^wa_w(w,y) dw\,\), provides us with the function
$A(w,y)\equiv\exp{a(w,y)}$. The function $B(w,y)$ is then found from Eq.~(\ref{eq3}).

The various solutions we are finding contain a great deal of redundancy that can be
absorbed into coordinate transformations. A redefinition of the scale of the $x^\mu$
coordinates, $x^\mu\rightarrow\lambda x^\mu$ changes only $A\rightarrow\lambda^2A$,
$a_0 \rightarrow a_0+2\log{\lambda}$, $c_0 \rightarrow c_0+2\log{\lambda}$. This
freedom can be used to set $c_0=0$.

The remaining freedom in arbitrarily choosing the function $a_0(y)$ locally
corresponds to a coordinate transformation $y\rightarrow\overline{y}(y)$. This can be
seen from the fact that as long as $a_0^\prime(y)\neq0$, the unique choice
$\overline{y}(y)=A(0,y)$ provides $\overline{A}(0,\overline{y})=\overline{y}$ and
the solution has no more continuous parameters left. 

Instead of the above choice of the $\overline{y}$ coordinate, however, we find it
more instructive to use coordinates in which the induced metric is conformally flat,
i.e. $A(0,y)=B(0,y)$. Such coordinates always exist and can be found from any
solution by the coordinate transformation to $\overline{y}=\int^y\sqrt{B/A}\,dy$.
Now expressing $a_y$ from Eq.~(\ref{eq:3'}) and using the definition of $c_0$
together with Eq.~(\ref{eq:c1}) we find (for w=0)
\begin{equation}
A_y=\pm\sqrt{\frac{2-5\sigma^2}{5}A^3+\frac{2}{15}\eta\,e^{\frac{5}{2}c_0}\sqrt{A}}.
\end{equation}
Using the variable $u(y)=u_0/\sqrt{A(0,y)}$ with
$u_0=-\eta\frac{exp{\frac{c_0}{2}}}{\sqrt[5]{\frac{3}{2}(2-5\sigma^2)}}$ we write
this in the form
\begin{equation}\label{eq:u}
\left(\frac{du}{dy}\right)^2=\left(\frac{2-5\sigma^2}{20}u_0^2\right)
\left(1-u^5\right).
\end{equation}
Note that, only on the brane,\footnote{The general relationship is
\(1-\frac{1}{v}=\left(\frac{u_0^2}A\right)^\frac52\frac{2-5\sigma^2}{2-5a_w^2}\).}
the variables $u$ and $v$ are related by $1-\frac{1}{v}=u^5$.

\subsection{Classification of the solutions}

\begin{figure}[htb]
\begin{center}
\fbox{\epsfxsize=12cm\epsfbox{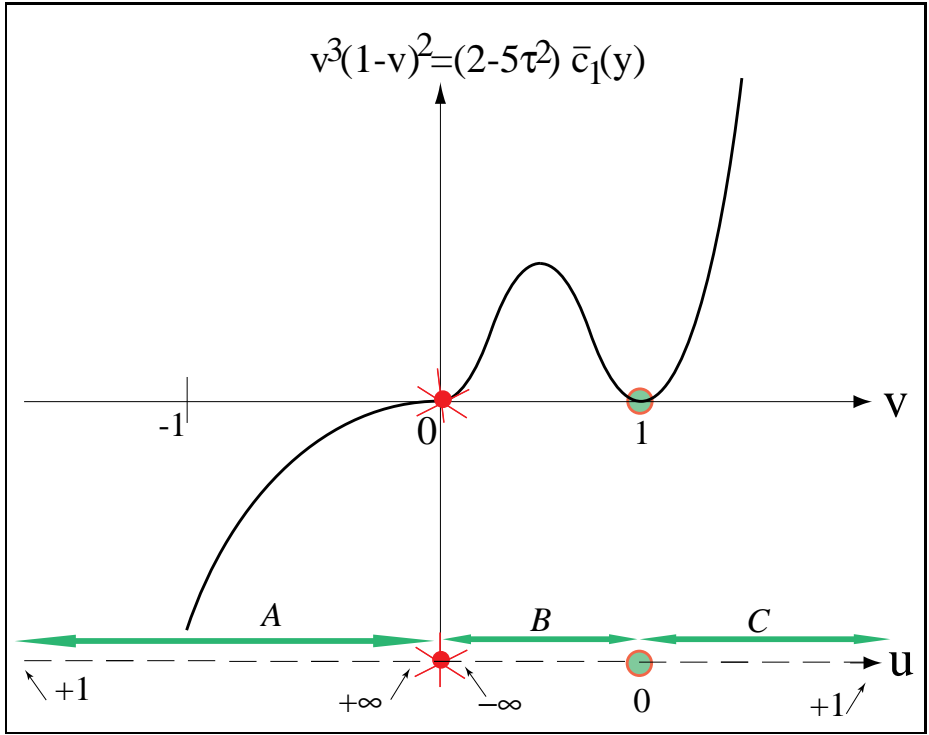}}
\end{center}
\caption{\protect\label{fig:v}The algebraic equation whose solution determines the
metric. The simultaneous change of the variable $u$, defined on the brane only, is
shown, together with the various cases discussed in the text. $v=0$ is a curvature
singularity and $v=1$ is an AdS horizon.}
\end{figure}

\begin{figure}[hbt]
\begin{center}
\fbox{\epsfxsize=16cm\epsfbox{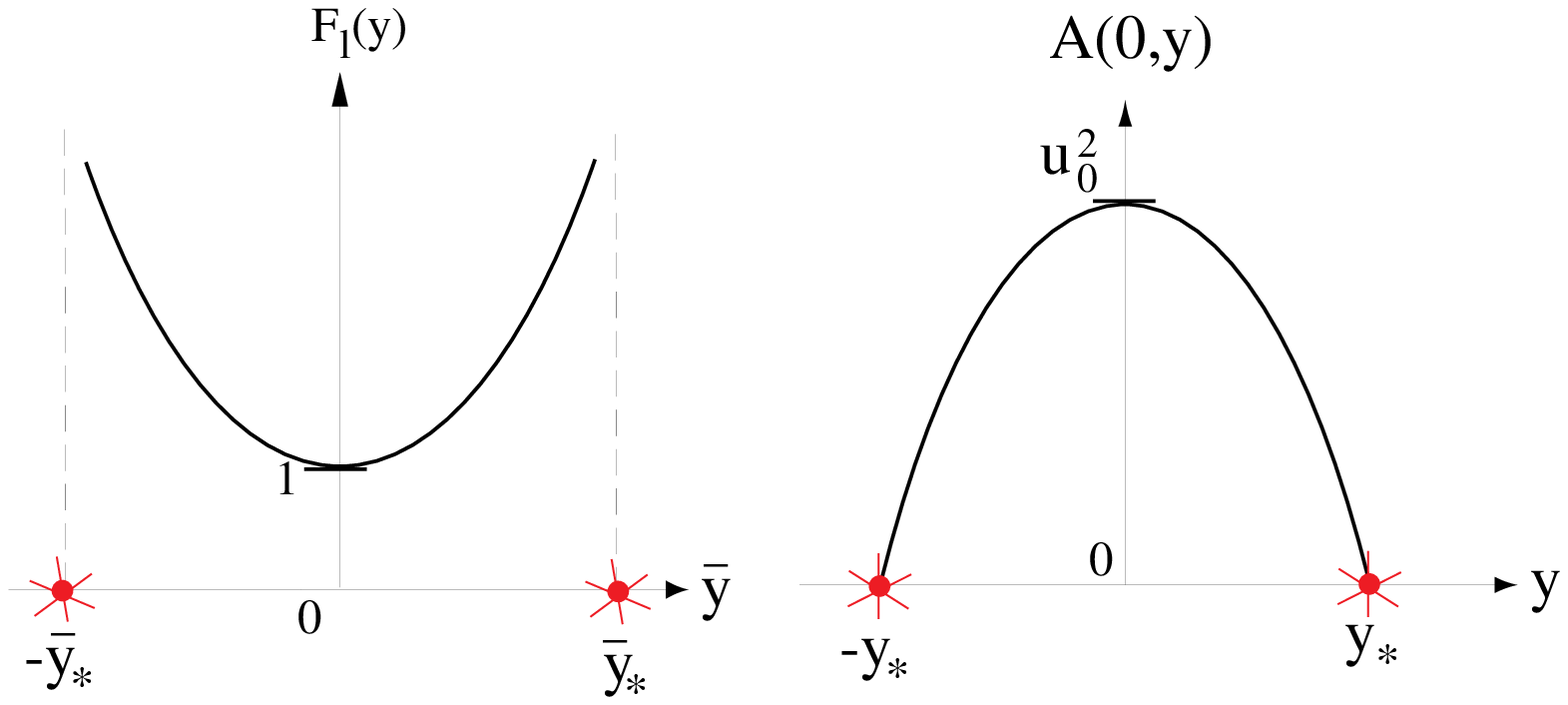}}
\end{center}
\caption{\protect\label{fig:Fs}The special function $F_s(\overline{y})$ is defined
for $-\overline{y}_*<\overline{y}<\overline{y}_*$, $\overline{y}_*\approx0.911$.
Also plotted is the conformal factor A(0,y) of the induced metric in case $A$,
$ds^2=A(0,y)\left[dy^2+\eta_\mu\nu dx^\mu dx^\nu\right]$. The points $y=\pm y_*$
correspond to curvature singularities.}
\end{figure}

The various cases can be best understood by following the change of $v$ in
Fig.~\ref{fig:v} as we move along the brane. In the case $A$ when $\sigma^2>\frac25$,
only $u>1$, i.e.~$v<0$ can satisfy Eq.~(\ref{eq:u}). It has one solution, unique up
to a translation of the $y$-coordinate,
\begin{equation}
u(y)=F_s\left[y\times u_0\sqrt{\frac{5\sigma^2-2}{20}}\right]
\mbox{\ \ \ i.e.\ \ \ }
A(0,y)=\frac{u_0^2}{F_s^2\left[y\times u_0\sqrt{\frac{5\sigma^2-2}{20}}\right]},
\end{equation}
where the special function $F_s(\overline{y})$ is defined as the solution of
\(|\overline{y}|=\int_0^{F_s}\frac{du}{\sqrt{u^5-1}}\), see Fig.~\ref{fig:Fs}.
The conformal factor of the induced metric, A(0,y) has a maximum at $y=0$ and
vanishes at $y=\pm y_*=\pm\overline{y}_*u_0\sqrt{\frac{5\sigma^2-2}{20}}$. By direct
calculation of the curvature of the induced metric one can establish the asymptotics,
\(R[g_{ab}^{(5)}]\propto\left(y_*\mp y\right)^{-\frac{10}{3}}\). The induced metric
has curvature singularities at these points which are easily seen to be at finite
proper distance.

\begin{figure}[hbt]
\begin{center}
\fbox{\epsfxsize=16cm\epsfbox{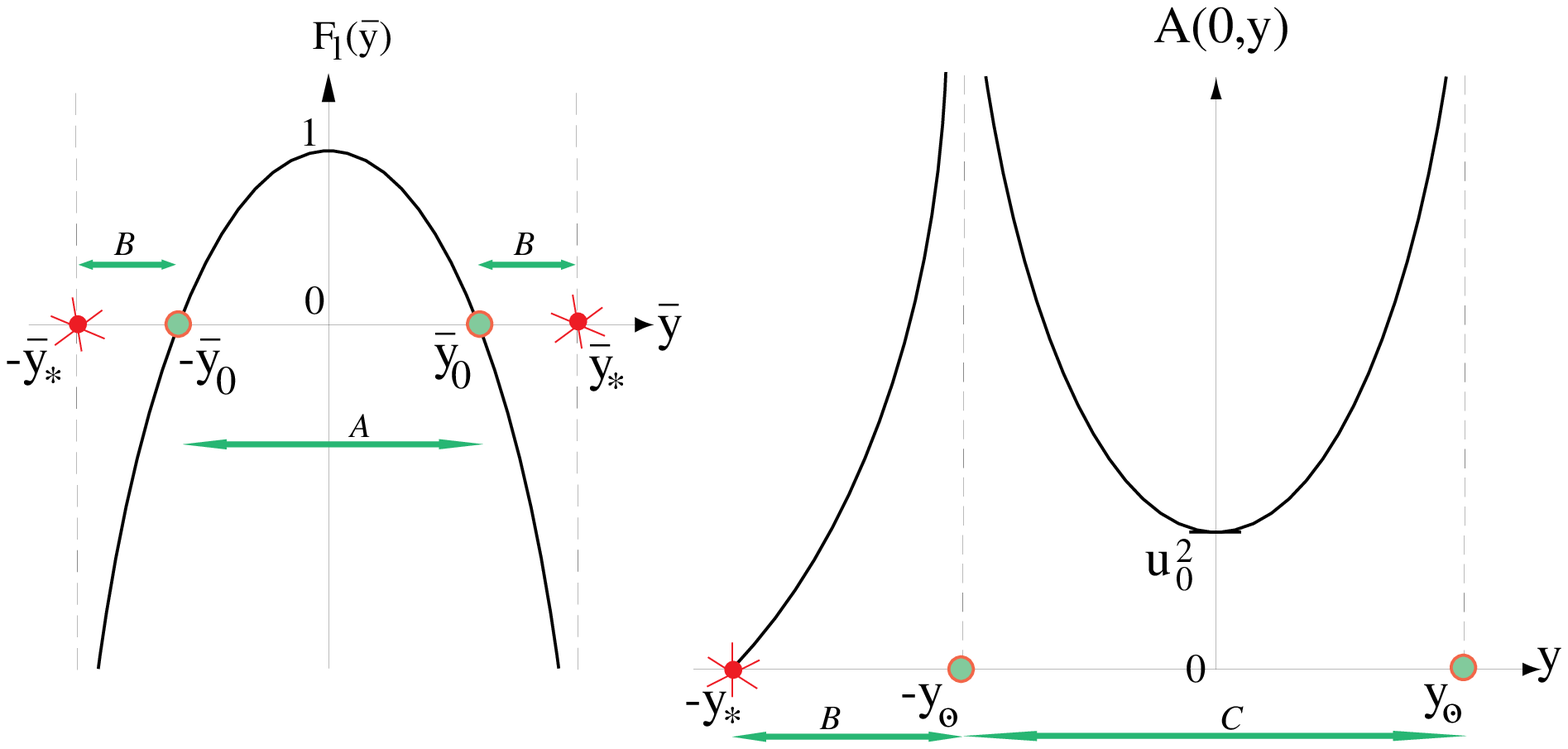}}
\end{center}
\caption{\protect\label{fig:Fl}The special function $F_l(\overline{y})$ is defined
for $-\overline{y}_*<\overline{y}<\overline{y}_*$, $\overline{y}_*\approx2.804$. It
is analytic at $\overline{y}=0$. Also plotted is the conformal factor of the induced
metric in cases $B,C$. The points $y=\pm y_*$ correspond to curvature singularities,
while $y=\pm y_\odot\,(\overline y_\odot\approx1.254)$ are infinitely far away and
represent $AdS$ horizons. The two regions with negative $F_l$ are related by a
$y\rightarrow-y$ coordinate transformation.}
\end{figure}

When $\sigma^2<\frac25$ we must have $u<1$, i.e. $v>0$ to satisfy Eq.~(\ref{eq:u}).
Again, we the solution can be written in terms of a special function as
\begin{equation}
u(y)=F_l\left[y\times u_0\sqrt{\frac{2-5\sigma^2}{20}}\right]
\mbox{\ \ \ i.e.\ \ \ }
A(0,y)=\frac{u_0^2}{F_l^2\left[y\times u_0\sqrt{\frac{2-5\sigma^2}{20}}\right]},
\end{equation}
where the special function $F_l(\overline{y})$ is now defined as the solution of
\(|\overline{y}|=\int_{F_l}^1\frac{du}{\sqrt{1-u^5}}\), plotted in
Fig.~\ref{fig:Fl}. In the two equivalent regions where $F_l<0$, case~$B$, $u$ is
negative and $0<v<1$. The conformal factor of the induced metric $A(0,y)$ tends to
zero at a curvature singularity at $y\rightarrow -y_*$, located at finite proper
distance, and the scalar curvature diverges as
\(R[g_{ab}^{(5)}]\propto\left(y-y_*\right)^{-5}\).
The function $A(0,y)$ increases with $y$ and at $y\rightarrow-y_\odot$ diverges. The
asymptotics is $A(0,y)\approx\frac{20}{2-5\sigma^2}\frac{1}{(y+y_\odot)^2}$,
corresponding to an asymptotically $AdS_5$ induced metric. Consequently, the point
$y=y_\odot$ is at infinite proper distance away, and the regions $B$ and $C$ should
be understood as different solutions. In the region $-y_\odot<y<y_\odot$, case $C$,
the conformal factor has a minimum and at the edges it diverges. The solution in
this region sees $AdS$ horizons in both directions. 

The definition of $c_0$ according to Eq.~(\ref{eq:c0}) supposes that the solution is
{\it not} maximally symmetric, $2-5a_w^2-10a_{ww}\neq0$. If it {\it is},
corresponding to $c_0\rightarrow-\infty$, we have a fourth type of
solution~\cite{KaloperBent,ANelson}. This case $D$ corresponds to $v(w,y)\equiv1$ as
one can see from the relation
\(\frac1v-1=\frac23\frac{2-5a_w^2-10a_{ww}}{2-5a_w^2}\rightarrow0\). In terms of
Fig.~\ref{fig:v}, the solution is ``sitting" in the $AdS$ point. In our coordinates
its form is
\begin{equation}
ds^2=dw^2+\frac{2-5\sigma^2}{2}\cosh^2\left[
\frac{w}{\sqrt{10}}+\mbox{arctanh}\left(\sqrt{\frac52}\sigma\right)\right]\times
\left\{dy^2+\eta_{\mu\nu}dx^\mu dx^\nu e^{y\sqrt{\frac{2-5\sigma^2}5}} \right\}.
\end{equation}
The induced metric on the brane is obviously $AdS_5$, as it should be.

The solutions we were finding for fixed values of $\sigma$ represent 6-dimensional
spacetimes with an embedded 4-brane. They have no parameters (other than $\sigma$)
that could not be absorbed into coordinate redefinitions. We will see in the next
section, that different values of $\sigma$ actually correspond to embedding branes
with different tensions in the same bulk metric and there are only three solutions
altogether, corresponding to the cases $A,B,C$ above. No 4-branes can be embedded
(within the confines of the metric Ansatz of Eq.~(\ref{eq:ansatz})) in any other
6-dimensional spacetime.\footnote{Excluding, of course, the case of a ``fine tuned"
brane, $\sigma^2=\frac{2}{5}$, that we do not consider here.}

In the exceptional case when $A(0,y)=const.$ along the brane it is not hard to see
that there is no solution to our equations.

\subsection{A note on localization of gravity to the brane}\label{subsec:localiz}

In the brane GNC's we are using, the large-$w$ asymptotic behavior of the functions
$A$ and $B$ can be better understood in the form
\begin{equation}\label{blowup}
A(w,y)\propto
B(w,y)=\left(\frac{1-\sqrt{\frac52}\sigma}2\right)^2e^{\sqrt{\frac25}\,w}
+\frac12\left(1-\frac52\sigma^2\right)
+\left(\frac{1+\sqrt{\frac52}\sigma}2\right)^2e^{-\sqrt{\frac25}\,w}.
\end{equation}
Far from the brane the first term dominates and evidently deconfines gravity. The
form of the metric in the Randall-Sundrum model~\cite{RS2} is
\begin{equation}
ds^2=dw^2-\eta_{\mu\nu}dx^\mu dx^\nu e^{-2kw},
\end{equation}
and the quick fall-off of the ``warp factor" $e^{-2kw}$ is essential in localizing
gravity to the brane (at least in the limit where the Kaluza-Klein excited modes can
be neglected). What is happening?

The resolution of the paradox hinges on two points. First, observe that a change in
$\sigma$ corresponds to a shift in $w$, in addition to rescaling $y$ and $x^\mu$, so
that a brane with a different tension could be placed parallel to the original one
in the same bulk. The RS brane corresponds to $\sigma^2\rightarrow\frac25$, i.e.
it is infinitely far away in $w$. The $\sigma\rightarrow\sigma_c$ limit is not
smooth.\footnote{That the limit is not smooth can also be seen in conformal
coordinates, $\rho>0$, $-\infty<z<+\infty$,
\(ds^2=\frac{10}{\rho^2}\left(\eta_{\mu\nu}dx^\mu dx^\nu+dz^2+d\rho^2\right)\). The
original $w$ coordinate corresponds to the new angular coordinate. The branes are
radial straight half lines in the $(z,\rho)$ plane while the fine tuned brane is
along $\rho=0,z>0$. The limit is not smooth because the metric diverges as
$\rho\rightarrow0$.} Now consider a brane with tension just smaller than the fine
tuned value,
$\sigma_c-\sigma\ll\sigma_c$. The behavior of the functions $A,B$ for increasing
$w$ is first governed by the second term, the warp factor falls off and there is
brane-localized gravity. By the time the exponential compensates for the smallness
of its coefficient and the first two terms start to dominate, the warp factor is
already ${\cal O}\left(\frac{1-\sigma/\sigma_c}{2}\right)\ll1$. It will grow again
exponentially. 

Second, the argument in~\cite{RS2} that leads to localization involves a Kaluza-Klein
{\it reduction}, i.e. dropping the brane-orthogonal components of the metric (such as
$g_{ww},g_{w\mu}$), in addition to the Kaluza-Klein {\it expansion} of the
brane-tangential modes (i.e. $g_{\mu\nu}$). While close to the brane the separation
of modes into brane-orthogonal and brane-tangential makes sense, far away it has no
covariant meaning. The actual separation and large-$w$ behavior depends on the
details of the model that explains the Kaluza-Klein reduction and cannot be expected
to be faithfully represented by the general framework of~\cite{RS2}. One may
conjecture by the same token that the increase in the warp factor is unphysical and
should be thrown away.

In any case the conclusion remains that in any scenario where the brane tensions are
not close to their fine tuned values does not automatically provide localized
gravity. In order to make a coordinate invariant statement one needs to restrict to
such coordinates in which the thrown-away modes ($g_{ww},g_{w\mu}$) are massive. It
cannot be decided without the explicit identification of these modes whether the
warp factor along the resulting $y=const$ lines shows the same behavior as
Eq.~(\ref{blowup}).
{\it This is equally relevant for codimension one branes in five dimensions.}

In this respect our most interesting solution, $C$, which has no curvature
singularities, can be seen to have the asymptotic behavior
\begin{equation}
A(w,y)\approx function(y)\times
e^{\sqrt{\frac25}|w|}\left(1+2\frac{1+\sqrt{\frac52}\sigma}{1-\sqrt{\frac52}\sigma}
e^{-\sqrt{\frac25}|w|}+
{\cal O}\left(e^{-\sqrt{\frac25}|w|}\right)^2\right),
\end{equation}
the same as the asymptotics of the known $AdS$ solutions.

\section{Intersecting branes}\label{sec:branes}

In the previous section we found all possible embeddings in GNC of the 4-brane. This
still does not answer the question whether another brane can be inserted in the same
bulk because for intersecting branes the two GNC's certainly do not co\"\i ncide. In
the CNC of one brane the other will not in general be straight. We were not able to
solve the equations for the branes in closed form, but we could (almost) show their
existence/unicity, and determine the relationship between the angle of intersection
and the brane tensions. 

\begin{figure}[htb]
\begin{center}
\fbox{\epsfxsize=9cm\epsfbox{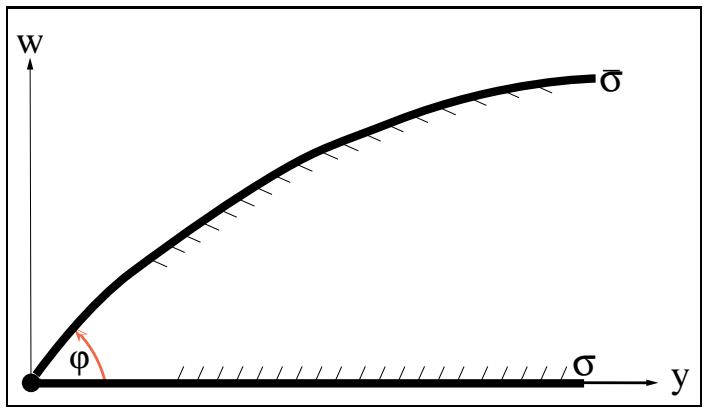}}
\end{center}
\caption{\protect\label{fig:brane}Cutting off a wedge with angle $\phi$. The
vertical lines are the geodesics orthogonal to the $\sigma$ brane.}
\end{figure}

We start with one solutions which permits a 4-brane with tension~$\sigma$ located at
$w=0$, keeping the $W>0$ half brane. If effect we are considering a half-brane at
$y\geq y_0$, and looking for another half-brane with tension $\tilde\sigma$,
starting off at some angle $0<\phi<+\infty$, and keeping a wedge between them as
shown in Fig.~\ref{fig:brane} (the wedge metric will of course be $2\pi$-periodic if
$\phi>2\pi$.)

The condition for the stability of the $\tilde\sigma$ brane is again the Israel
condition $K_{ab}=+\frac{\tilde\sigma}{2}h_{ab}$, where the opposite sign is due to the fact that
this time we are keeping the bulk to the opposite side of the brane. We describe the
brane with a pair of functions
\begin{equation}
\tilde w(\lambda), \tilde y(\lambda) \mbox{\ \ \ with\ \ \ }
 \tilde w(0)=0, \tilde y(0)=y_0\mbox{\ \ \ and\ \ \ }0\leq\lambda<\infty.
\end{equation}

The Israel conditions become
\begin{eqnarray}\label{eq:b1}
2\left(\ddot{\tilde y}\dot{\tilde w}-\ddot{\tilde w}\dot{\tilde y}\right)
+b_y\,\dot{\tilde w}\dot{\tilde y}^2
+b_w\,\dot{\tilde w}\dot{\tilde y}\,\left(2\dot{\tilde w}+B\dot{\tilde y}\right)
&=&\frac{\tilde\sigma}{\sqrt B}\\
\frac{a_y}{\sqrt B}\,\dot{\tilde w}-a_w\sqrt B\,\dot{\tilde y}
&=&-\tilde\sigma\label{eq:b2}
\end{eqnarray}
with $B=e^b$, where we fixed the $\lambda$ variable by requiring
\begin{equation}\label{eq:lambda}
{\dot{\tilde w}}^2+B{\dot{\tilde y}}^2=1.
\end{equation}

Now Eq.~(\ref{eq:b2}) can be viewed as a first order differential equation for the
function $\tilde w(\tilde y)$ with initial conditions set by $\tilde w(y_0)=0$ and by
Eq.~(\ref{eq:lambda}). Up to a discreet ambiguity, this system has a unique solution.
We could not show in complete generality that Eq.~(\ref{eq:b1}) is a consequence of
Eq.~(\ref{eq:b2}), we expanded the solution of Eq.~(\ref{eq:b2}) in a power series
around $\lambda=0$ keeping terms up to ${\cal O}(\lambda^5)$ and saw that they did
indeed satisfy Eq.~(\ref{eq:b1}), up to ${\cal O}(\lambda^3)$. (The loss of two
powers of $\lambda$ is due to the three derivatives figuring in Eq.~(\ref{eq:b1}).)
We take this as an indication that Eq.~(\ref{eq:b1}) is indeed not an independent
equation.

We can define the angle between the in a covariant way by requiring
$\cos\phi=n_ag^{ab}\tilde n_b$, where $n_a$ and $\tilde n_b$ are the unit normals of
the branes at the intersection. In our coordinates we find \(
\left(\sin\phi|\cos\phi\right)=
\left(\dot{\tilde w}\left|\sqrt B\dot{\tilde y}\right.\right)_{\lambda=0}.
\)
Then, substituting this definition into Eq.~(\ref{eq:b2}) we see that the brane
tension and the position of the junction (i.e. $y_0$) determines the angle:
\begin{equation}\label{eq:angle}
\frac{a_y}{\sqrt B}\sin\phi+\sigma\cos\phi+\tilde\sigma=0.
\end{equation}
A trivial check of this calculation is that $\phi=\pi$, $\tilde\sigma=\sigma$ is
always a solution: the brane can always be continued across the point $y_0=0$. We
also see that once Eq.~(\ref{eq:angle}) is satisfied, there is one unique solution
to Eqs.~(\ref{eq:b1}-\ref{eq:lambda}).

Before discussing the solutions to Eq.~\ref{eq:angle} we look at the induced
metric on the $\tilde\sigma$ brane. A simple substitution shows
\(d\tilde s^2=d\lambda^2+A\eta_{\mu\nu}dx^\mu dx^\nu\), which can be written in a
conformally flat from by introducing the new variable
$Y=\int_0^\lambda\frac{d\lambda}{\sqrt A}$ along the new brane:
\(d\tilde s^2=A\left(dY^2+\eta_{\mu\nu}dx^\mu dx^\nu\right)\). Note that the $x^\mu$
coordinates did not have to be rescaled, and the conformal factor A is the same at
the junction in the two induced metrics. The bulk metric can now be written in GNC's
of the $\tilde\sigma$ brane in the form \(ds^2=dW^2+\tilde B(W,Y)dY^2+\tilde
A(W,Y)\eta_{\mu\nu}dx^\mu dx^\nu\); we know $\tilde A(0,Y)=\tilde B(0,Y)=A(\tilde
w,\tilde y)$. But this must be one of our previously found solutions which comes
with its $\tilde v(W,Y)$ function. A direct calculation of $\tilde v(0,0)$ gives,
using Eq.~(\ref{eq:angle}), and observing $v=\frac{2-5\sigma^2}{5}\,\frac{\tilde
B}{\tilde a_Y^2}$
\begin{equation}
\tilde V\equiv\tilde\sigma^2+\frac{2-5\tilde\sigma^2}{5\,\tilde v(0,0)}\ \ =\ \ 
V\equiv\sigma^2+\frac{2-5\sigma^2}{5\,v(0,0)}.
\end{equation}
In other words, the quantity $V$ must be the same on the two sides of the wedge.

It will prove important to note now that once $\tilde\sigma,\tilde v(0,0)$ and
$\tilde A(0,0)$ are specified as they are, there are only two possible induced
metrics on the new brane. The ambiguity results of the two ways to invert the
functions $F_{l,r}$ and can be resolved by also specifying the sign
$\tilde\kappa=\mbox{sign}\,{\tilde a_Y(0,0)}=\pm1$. Then we will be able to argue
that any two induced metrics with the same $\sigma,v(0,0),A(0,0)$ and $\kappa$
co\"\i ncide. This will be the situation when a number of wedges is pasted together
around one 3-brane.

\begin{figure}[htb]
\begin{center}
\fbox{\epsfxsize=9cm\epsfbox{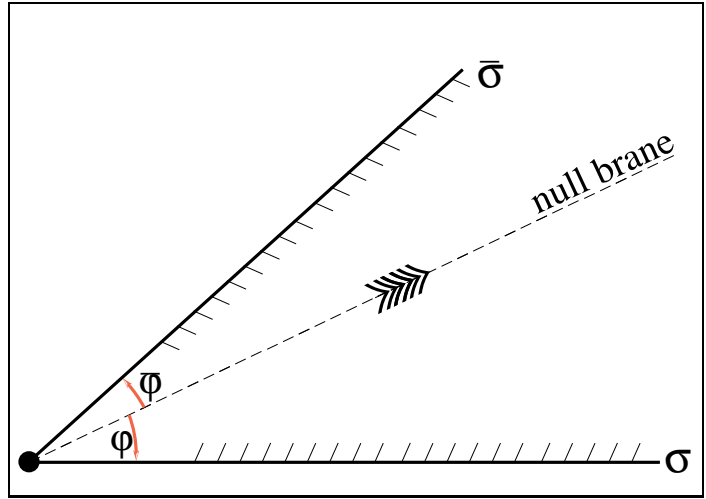}}
\end{center}
\caption{\protect\label{fig:wedge}Cutting out a wedge between two branes with
tensions $\sigma$ and $\tilde\sigma$. The double arrow shows the direction in A
increases.}
\end{figure}

The solutions of Eq.~(\ref{eq:angle}) can be visualized by the following auxiliary
device. It is always possible to find a unique solution $\phi=-\varphi$ to
Eq.~(\ref{eq:angle}) with $\tilde\sigma=0$, also requiring $\tilde\kappa=+1.$ This
corresponds to a ``null brane" represented in Fig.~\ref{fig:wedge}. It is not hard
to see that we could have started with the null brane and looked for the angles at
which the two $\sigma,\tilde\sigma$ branes to cut out the wedge. The equations for
their angles become
\begin{equation}
\frac{a_y}{\sqrt B}\sin\tilde\varphi+\tilde\sigma=0 \mbox{\ \ \ and\ \ \ }
\frac{a_y}{\sqrt B}\sin\varphi+\sigma=0,
\end{equation}
where $\frac{a_y}{\sqrt B}=\sqrt{V}$ now refers to the null brane. The solutions
that give the correct sign for the $a_y$'s are, with
$-\frac\pi2\leq\arcsin\leq\frac\pi2$,
\begin{equation}\label{eq:anglesoln}
\varphi=\pi\frac{1-\kappa}2-\kappa\arcsin\frac\sigma{\sqrt V}+2\pi N
\mbox{\ \ \ and\ \ \ }
\tilde\varphi=
 \pi\frac{1-\tilde\kappa}2-\tilde\kappa\arcsin\frac{\tilde\sigma}{\sqrt V}+
 2\pi\tilde N.
\end{equation}

Based on the ``compatibility" of different branes, the possible setups can be
classified as follows. It is not possible to move along any brane from one category
to another.
\begin{itemize}
\item{$0<V<\frac25$, only branes with below critical tension, $\sigma^2<\frac25$ can
be used, all type $C$ (i.e. $v>1$) with no curvature singularity. The angles are
determined by the tensions and the parameter $V$ which is fixed in any one junction
but varies from junction to junction}
\item{$V=\frac25$, the spacetime is $AdS_6$, only type $D$ branes with below critical
tension, $\sigma^2<\frac25$ can be used. The angles are determined by the tensions
and there is no additional parameter.}
\item{$V>\frac25$, one can accommodate both type $B$ branes with $0<\sigma^2<\frac25$
and type $A$ branes with $\sigma^2>\frac25$. There is again a parameter $V$ in the
junctions. All induced metrics contain curvature singularities.}
\end{itemize}

\section{Sawing together the wedges}\label{sec:model}

Any combination of intersecting branes with angles between them set according to
Eq.~(\ref{eq:anglesoln}) represents a stationary point of the action in
Eq.~(\ref{eq:action}). The $V$ are parameters of the solution, one at each
intersection. We must make sure, however, that a globally defined metric exists when
the pieces we have discussed are sewn together.

Each brane carries orbifold boundary conditions, which are easiest to state in
GNC's: we identify points with the sign of the corresponding coordinate $w$ flipped.
This allows at most two different sets of 4-brane tensions ($\sigma$) in each
intersection: every second one most be equal. The corresponding signs $\kappa$
should also be chosen equal. In we join an odd number of branes in one intersection,
then all must have the same $\sigma$ and $\kappa$.

When $n=1,2,\dots$ branes intersect along one 3-brane, the $n+1^{th}$ one must be
identified with the first one. There are two conditions that must be met for
consistency. (i) The induced metric on the 3-brane must be the same for all wedges.
This condition is always satisfied because the functions $A(0,0)$ are always the
same and no rescaling of the $x^\mu$ coordinates was necessary. (ii) The induced
metric on the $n+1^{th}$ brane must be identical with the induced metric on the
$1^{st}$. This condition is again automatically satisfied because, as we saw, the
equality of $V,A,\kappa$ and $\sigma$ is sufficient for that. 

\begin{figure}[htb]
\begin{center}
\fbox{\epsfxsize=16cm\epsfbox{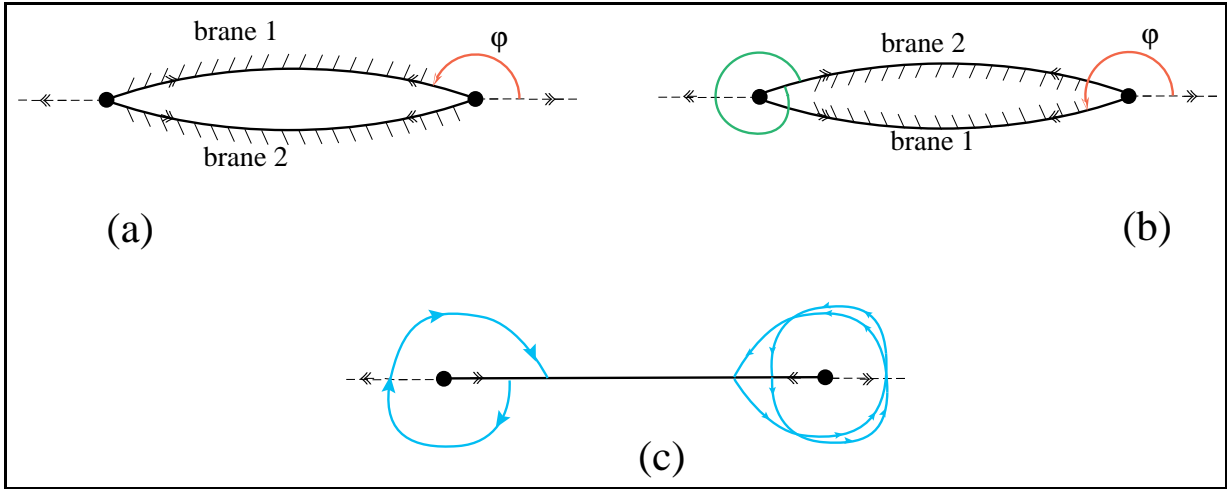}}
\end{center}
\caption{\protect\label{fig:strip}The illustrative geometry of the brane strip.
In~(a), the angles are distorted so that in actuality $\varphi>180^0$, while in (b)
the angles are correct but the bulk is shown partially overlapping. The spirals show
the part of the bulk that is retained. Finally, the two branes are identified as
in~(c). The double arrows show the direction in which $A$ increases. In (c) some of
the brane orthogonal geodesics are shown (these represent the $y=const.$ lines in
brane CNC).}
\end{figure}

\begin{figure}[htb]
\begin{center}
\fbox{\epsfxsize=9cm\epsfbox{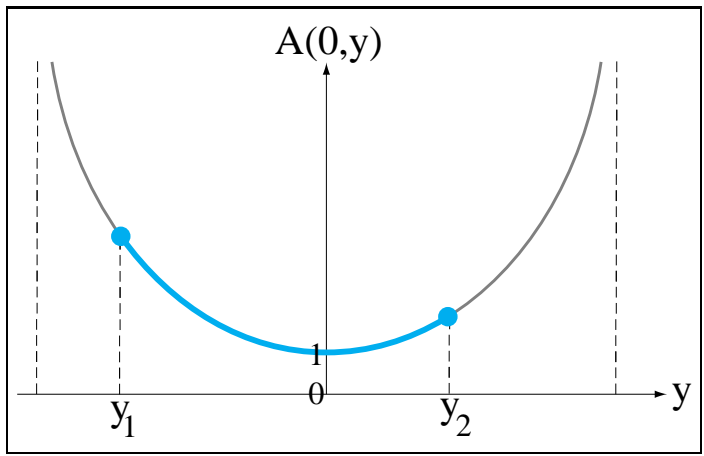}}
\end{center}
\caption{\protect\label{fig:conf} The conformal factor along the brane strip.}
\end{figure}

In the following we consider a case with one positive tension 4-brane with two
junctions around which the brane is identified with its own reverse side, shown in
Fig.~\ref{fig:strip}. We require, due to the effectively two dimensional gravity on
the $w,y$ plane, an excess angle $0<\lambda<\pi$. This is possible only with
$\kappa=-1$; and in order to have no curvature singularities, we will use type $C$
branes. Because then $a_0(y)$ must increase towards both ``junctions", it must have a
minimum (at $y=0$) on the brane between then. Using the excess angles $\lambda_{1,2}$
as parameters, Eq.~(\ref{eq:anglesoln}) tells the parameter (on each end separately)
$V=\frac{\sigma^2}{\sin^2\frac\lambda2}$.
Because for the $C$ type branes $V<\frac25$, we can accommodate only an excess angle
at least as large as $\sin\frac{\lambda}2>\sqrt{\frac52}\sigma$.
The 3-brane tension cannot be much smaller than the 4-brane tension. In brane GNC's
the conformal factor is $A(0,y)=F_l^{-2}\left[y\sqrt{\frac{2-5\sigma^2}{20}}\right]$,
shown in Fig.~\ref{fig:conf}. We calculated the position of the two ends of the
brane, with the result
\begin{equation}
y_j=\frac{(-1)^j}{\sqrt{\frac{2-5\sigma^2}{20}}}\int_{u_j}^1\frac{du}{\sqrt{1-u^5}} 
\mbox{\ \ \ with\ \ \ } j=1,2.
\end{equation}
The proper distance of the endpoints from the origin is
\begin{equation}\label{eq:proper}
l_j=\frac{(-1)^j}{\sqrt{\frac{2-5\sigma^2}{20}}}\int_{u_j}^1\frac{du}{u\sqrt{1-u^5}}
\mbox{\ \ \ with\ \ \ }
u_j=\left[\frac{2\sin^2\frac{\lambda_j}2-5\sigma^2}{(2-5\sigma^2)
\sin^2\frac{\lambda_j}2}\right]^\frac15.
\end{equation}
Consistency requires that this distance should be $l\gg1$ (otherwise the underlying
field theory description of gravity is not justified). This requires that the
4-brane tension not be much smaller than $\lambda$ and also either $2-5\sigma^2\ll1$
or $0<\sin\frac{\lambda}2-\sqrt{\frac52}\sigma\ll1$.

The presence of a vacuum energy density on the 3-brane requires the proper amount of
excess angle. When this relationship is not satisfied, our Ansatz does not give a
solution: the induced metric on the 3-brane is not flat any more. It is a trivial
exercise to derive the equations of motion when a maximally symmetric but not flat
4-dimensional metric is allowed, i.e. $-dt^2+d{\bf x}^2$ is replaced by
$-dt^2+d{\bf x}^2e^{2 h t}$ in Eq.~(\ref{eq:ansatz}). We find that
Eqs.~(\ref{eq4},\ref{eq:4'}) are not modified, so that the function $A(w,y)$ is not
modified. The modification occurs because Eq.~(\ref{eq:3'}) gets replaced by
\begin{equation}\label{eq:3'h}
B=\frac{3a_y^2}{2-5a_w^2-4a_{ww}+12\frac{h^2}{A}},
\end{equation}
so that $B(w,y)$ changes. This change also affects the intersection angles,
\begin{equation}
\frac{\tan^2\varphi_h}{\tan^2\varphi}=\frac{V}{V+4\frac{h^2}{A}}.
\end{equation}

If the ``mistuning" is little,
\(\delta\lambda=\lambda_h-\lambda=2(\phi_h-\phi)\ll1\),
the result is a small $h^2=-\frac{A V}{2\sin\lambda}\delta\lambda$. This translates
to a small observable Hubble constant (the induced metric must be brought to the
form $ds_4^2=-dt'^2+d{\bf x}'^2e^{2Ht'}$, $H=h/\sqrt A$),
\begin{equation}
H^2=-\frac{\sigma^2}{\sin^2\frac{\lambda}2}\times\frac{\delta\lambda}{2\sin\lambda}.
\end{equation}

Given that $\lambda$ cannot be made much smaller than $\sigma$, any sizable
``mistuning" of $\lambda$ would result in a Hubble constant on the 6-dimensional
Planck scale.

Up to this point however, we have not required any fine tuning. The 4-dimensional
vacuum density, through the topologically required conical singularity of the
metric enforces the adjustment of the parameter $V$ to its required value. However,
when a 3-brane tension is added to the action,
\begin{equation}\label{eq:fullaction}
S[g]=\int_{bulk}(\Lambda_6-2M_6^4 R[g])+\int_{4-brane(s)}\Lambda_5
+\int_{3-brane}\Lambda_4,
\end{equation}
the added term can be written as
\begin{equation}
\int_{3-brane}\Lambda_4=\Lambda_4\ A^2(0,0)\,\int_{3-brane}d^4x.
\end{equation}
In order to avoid that the spacetime collapses onto the point where $A(0,0)$ is
minimum, i.e. $y=0$, additional repulsive interactions have to be introduced. Their
strength has to be fine tuned to ensure that the equilibrium length of the 4-brane
is that of Eq.~(\ref{eq:proper}).

\section{Conclusions}

The main result of this paper is that there exists embeddings of 4-branes in
less than maximally symmetric 6-dimensional spacetimes that preserve 4-dimensional
Poincar\'e invariance. Then different points on the 4-brane are not physically
equivalent as they were in the $AdS$ case. The angle between intersecting branes,
in any fixed intersection point, is always determined by the nondynamical
parameters through the Israel conditions $K_{ab}=-\frac{\sigma}{2} h_{ab}$. We saw that there
is no obstacle to building a global metric around the intersection and the additional
parameter can be translated to an arbitrary excess angle. We also saw that
there is no solution (without fine tuning the 4-brane tensions) with constant
warp factor along the 4-brane. As a consequence, as we illustrated on a simple
example, a 3-brane with tension tries to ``move" to the mininum of that warp factor
and additional interactions are needed to stabilize the 3-brane. We note that in all
suggested solutions to the cosmological constant
problem~\cite{{Luty},{Arkani-Hamed:2000eg},{Kachru:2000hf}} in brane contexts some
sort of tuning of the bulk interactions has been required.

\section{Acknowledgments}

The author thanks Bob Holdom for many useful discussions and stressing the importance
of finding flat directions, as well as Hael Collins for discussions.


\thebibliography{99}
\bibitem{RS2}L.~Randall and R.~Sundrum,
``An alternative to compactification,''
Phys.\ Rev.\ Lett.\  {\bf 83} (1999) 4690
[hep-th/9906064].
\bibitem{Intersect1}
N.~Arkani-Hamed, S.~Dimopoulos, G.~Dvali and N.~Kaloper,
``Infinitely large new dimensions,''
Phys.\ Rev.\ Lett.\  {\bf 84}, 586 (2000)
[hep-th/9907209].
\bibitem{IntersectAll}
N.~Arkani-Hamed, S.~Dimopoulos, G.~Dvali and N.~Kaloper,
``Manyfold universe,''
hep-ph/9911386;
%
S.~Nam,
``Mass gap in the Kaluza-Klein spectrum in a network of branewords,''
[hep-th/9911237];
N.~Kaloper,
``Crystal manyfold universes in AdS space,''
Phys.\ Lett.\  {\bf B474}, 269 (2000)
[hep-th/9912125];
%
J.~Cline, C.~Grojean and G.~Servant,
``Inflating intersecting branes and remarks on the hierarchy problem,''
Phys.\ Lett.\  {\bf B472}, 302 (2000)
[hep-ph/9909496].
%
\bibitem{KaloperBent}N.~Kaloper,
``Bent domain walls as braneworlds,''
Phys.\ Rev.\  {\bf D60}, 123506 (1999)
[hep-th/9905210].
\bibitem{ANelson}A.~E.~Nelson,
``A new angle on intersecting branes in infinite extra dimensions,''
hep-th/9909001.
\bibitem{RS1}L.~Randall and R.~Sundrum,
``A large mass hierarchy from a small extra dimension,''
Phys.\ Rev.\ Lett.\  {\bf 83}, 3370 (1999)
[hep-ph/9905221].
\bibitem{Csaki}Cs.~Cs\'aki and Yu.~Shirman,
``Brane junctions in the Randall-Sundrum scenario,''
Phys.\ Rev.\  {\bf D61}, 024008 (2000)
[hep-th/9908186].
\bibitem{Israel}W.~Israel,
``Singular Hypersurfaces And Thin Shells In General Relativity,''
Nuovo Cim.\  {\bf B44 S10}, 1 (1966).
\bibitem{Chodos}A.~Chodos and E.~Poppitz,
``Warp factors and extended sources in two transverse dimensions,''
Phys.\ Lett.\  {\bf B471}, 119 (1999)
[hep-th/9909199].
\bibitem{Luty}J.~Chen, M.~A.~Luty and E.~Pont\'on,
``A critical cosmological constant from millimeter extra dimensions,''
hep-th/0003067.
\bibitem{Arkani-Hamed:2000eg}
N.~Arkani-Hamed, S.~Dimopoulos, N.~Kaloper and R.~Sundrum,
``A small cosmological constant from a large extra dimension,''
hep-th/0001197.
\bibitem{Kachru:2000hf}
S.~Kachru, M.~Schulz and E.~Silverstein,
``Self-tuning flat domain walls in 5d gravity and string theory,''
hep-th/0001206.
\end{document}